\documentclass[12pt]{article}
\usepackage{latexsym}
\usepackage{graphicx}
\usepackage{dcolumn}
\usepackage{bm}
\usepackage{amsmath}
\usepackage{amssymb}
\usepackage{color}
\usepackage{slashed}
\usepackage{authblk}
\title{The time distribution of quantum events}
\author{Danijel Jurman\footnote{e-mail: djurman@irb.hr} $\;$}
\author{Hrvoje Nikoli\'c\footnote{e-mail: hnikolic@irb.hr}}
\affil{Theoretical Physics Division, Rudjer Bo\v{s}kovi\'{c} Institute, \\
P.O.B. 180, HR-10002 Zagreb, Croatia}
\begin{document}
\maketitle
\begin{abstract}
We develop a general theory of the time distribution of quantum events,
applicable to a large class of problems 
such as arrival time, dwell time and tunneling time.
A stopwatch ticks until
an awaited event is detected, at which time the stopwatch stops.
The awaited event is represented by a projection operator $\pi$,
while the ideal stopwatch is modeled as a series of projective measurements 
at which the quantum state gets projected
with either $\bar{\pi}=1-\pi$ (when the awaited event does not happen) 
or $\pi$ (when the awaited event eventually happens).
In the approximation in which the time $\delta t$ between the subsequent measurements 
is sufficiently small (but not zero!), we find a fairly simple general formula 
for the time distribution ${\cal P}(t)$, representing the probability density that the 
awaited event will be detected at time $t$. 
\end{abstract}

\vspace*{0.5cm}

\maketitle

\section{Introduction} 

In the standard formulation of quantum mechanics (QM) time is not an operator \cite{pauli},  
which leads to various incarnations of the problem of time in QM \cite{muga1,muga2}.
From an operational point of view, particularly important is the large class of problems
in which one asks what is the probability that a given event will happen at time $t$.
Some members of this class of problems are the arrival time 
\cite{V1.10,V2.4,muga-physrep,rovelli,delgado,galapon2,anastopoulos2,halliwell1,anastopoulos,vona,dhar,halliwell2,galapon},
dwell time \cite{V2.5,kelkar,yearsley} 
and tunneling time \cite{V1.12,yan,ordonez,lunardi,delbarco,sokolowski,baytas}. 
Several inequivalent theoretical approaches have been proposed for each member 
in this class (see e.g. the review \cite{muga-physrep}) 
and there is no consensus which of those approaches, if any, should be the correct one. 
Moreover, although the different problems in this class are all related to each other, 
in the literature each of those problems is usually treated separately from the other problems. 
A satisfying general theory that treats all such problems on an equal footing seems to be missing.

In this paper we develop such a general theory. Our approach is strictly operational in the sense 
that we study the probability that an event will be {\em detected} at time $t$.
But in addition to being operational, our approach is also very general, in the sense that the theory does
not depend on details of the detector. All essential quantum ingredients of the theory 
are formulated in the Hilbert space ${\cal H}$ of the studied system, without a need to study
explicitly the Hilbert space of detector states.
(Nevertheless, the states of the detector can also be included in the description,
which we discuss too.)
With such an approach the detector is specified by only two quantities:
the time resolution $\delta t$ of the detector and the projector $\pi$ acting in ${\cal H}$ that represents the detected event.
A typical example useful to have in mind is a detector that
determines whether the particle has appeared inside the spatial region 
$V_{\rm det}$, in which case 
\begin{equation}\label{I1}
 \pi=\int_{V_{\rm det}} d^3x\, |{\bf x}\rangle\langle{\bf x}| ,
\end{equation}
where $|{\bf x}\rangle$ are position eigenstates of the considered particle.

In addition to $\pi$ and $\delta t$, our general theory involves also  
the initial state $|\psi_0\rangle$ and the intrinsic Hamiltonian $H$ 
of the studied system, where by ``intrinsic'' we mean that $H$ acts on the states in ${\cal H}$ 
and does not involve interaction with the detector.
In the absence of detector, the time evolution of the state is given by 
$|\psi(t)\rangle=e^{-iHt/\hbar}|\psi_0\rangle$. When the detector is working
and when $\delta t$ is sufficiently small 
(the precise meaning of ``sufficiently small'' will be specified later)
the main new result of this paper can be summarized by a concise formula for 
the probability density ${\cal P}(t)$ that the event will be detected at time $t$:
\begin{equation}\label{I2}
 {\cal P}(t)=w(t)e^{-\int_0^t dt'w(t')} ,
\end{equation}
where
\begin{equation}\label{I3}
 w(t)=\frac{1}{\delta t}\langle\psi_c(t)|\pi|\psi_c(t)\rangle ,
\end{equation}
\begin{equation}\label{I4}
 |\psi_c(t)\rangle=e^{-iH\delta t/\hbar}e^{-i\overline{H}(t-\delta t)/\hbar}|\psi_0\rangle ,
\end{equation}
\begin{equation}
 \overline{H}=\bar{\pi}H\bar{\pi}, \;\;\; \bar{\pi}=1-\pi .
\end{equation}
The derivation of these formulas and their physical meaning is explained in the rest of the paper.
We outline the final formulas above so that a reader more interested in applications than in 
abstract theory can, in principle, skip abstract theory presented in Secs.~\ref{SECmain} and \ref{SECnotes}
and jump to Secs.~\ref{SECdecay}, \ref{SECarrival} and \ref{SECdwelltunn} where we explain 
the general principles of how can those formulas be applied.

The paper is organized as follows. 
The main principles of the theory, giving rise to a derivation of the formulas above, are presented in Sec.~\ref{SECmain},
while various additional aspects of the theory that may be needed for a deeper conceptual understanding 
are discussed in Sec.~\ref{SECnotes}. 
After that, 
we outline general principles of how to apply the theory to a decay of an unstable state in Sec.~\ref{SECdecay},
to arrival time in Sec.~\ref{SECarrival}, and to dwell time and tunneling time in Sec.~\ref{SECdwelltunn}. 
The conclusions are drawn in Sec.~\ref{SECconclusion}.

\section{Derivation of the main formula}
\label{SECmain}

Suppose that initially, at time $t_0=0$, a quantum system is prepared in the state $|\psi_0\rangle$. 
Let $H$ be the intrinsic Hamiltonian of the system so that, in the absence of detection,
the state of the system evolves as
\begin{equation}\label{Hevol} 
 |\psi(t)\rangle=e^{-iHt}|\psi_0\rangle .
\end{equation}
(To save writing, unless specified otherwise we work in units $\hbar=1$.)

Now consider a stopwatch with a time resolution $\delta t$, so that it ticks at times 
$t_1=\delta t$, $t_2=2\delta t$, $t_3=3\delta t$, etc. Furthermore, suppose that the stopwatch is coupled to 
a detector, so that, at 
each $t_k$, the detector checks whether the system has an awaited property defined by a projector $\pi$.
If the awaited property is detected at a given time $t_k$, then the effect of the detector is to induce the 
``wave function collapse''
\begin{equation}\label{collapse1}
|\psi\rangle \rightarrow \frac{\pi |\psi\rangle}{|\pi |\psi\rangle|} .
\end{equation}
At that time, the stopwatch stops and the experiment is over. The final state 
of the stopwatch (e.g. the final spatial position of the clock's needle) records the time $t_k$ of detection.
Our goal is to determine the probability $P(t_k)$ 
that the experiment will be over at the time $t_k$.

From a practical point of view, the collapse corresponds to a gain of new information,
that is, the information that the property has been detected at $t_k$.
Our analysis will not depend on whether the collapse is interpreted as a real physical event, or just as an
update of information. The equations that we shall write will not depend on the interpretation of QM.
(But some notes on interpretations are given in Sec.~\ref{SECunitary}.)

Furthermore, we assume that the detector has a perfect efficiency. Hence, if the awaited property is not detected
at $t_k$, then we also gain a new information - the information that the system does not have the property $\pi$
at time $t_k$. Hence an absence of detection also induces a ``wave function collapse'', namely
\begin{equation}\label{collapse2}
|\psi\rangle \rightarrow \frac{\bar{\pi} |\psi\rangle}{|\bar{\pi} |\psi\rangle|} ,
\end{equation}
where
\begin{equation}
 \bar{\pi}=1-\pi .
\end{equation}

For example, suppose that the event has not been detected at $t_1$. Then the state evolution from $t_0$ to $t_1$ is
\begin{equation}
 |\psi_0\rangle \rightarrow e^{-iH\delta t} |\psi_0\rangle \rightarrow 
\frac{\bar{\pi} e^{-iH\delta t} |\psi_0\rangle}{| \bar{\pi} e^{-iH\delta t}|\psi_0\rangle|}
= \frac{V|\psi_0\rangle}{|V|\psi_0\rangle|} ,
\end{equation}
where
\begin{equation}
 V\equiv \bar{\pi} e^{-iH\delta t} .
\end{equation}
Likewise, if the event has not been detected at $t_1$ and $t_2$, then the state evolution from $t_0$ to $t_1$, 
and then from $t_1$ to $t_2$, can be written more succinctly as 
\begin{equation}
 |\psi_0\rangle \rightarrow \frac{V|\psi_0\rangle}{|V|\psi_0\rangle|} \rightarrow  \frac{VV|\psi_0\rangle}{|VV|\psi_0\rangle|} .
\end{equation}
By induction, if the event has not been detected at all times from $t_1$ to $t_k$, 
then then the state evolution from $t_0$ to $t_k$ can be written compactly as
\begin{equation}\label{colV^k}
 |\psi_0\rangle \rightarrow \frac{V^k|\psi_0\rangle}{|V^k|\psi_0\rangle|} .
\end{equation}

Now suppose that $\delta t$ is a sufficiently short time, so that the state is not changed much during  $\delta t$
by the evolution governed by $H$. This means that $|\langle\psi|e^{-iH\delta t}|\psi\rangle|=1-\epsilon$, with
$\epsilon\ll 1$. Then $V^k$ can be approximated as
\begin{eqnarray}\label{V^k_first}
V^k & = & \bar{\pi} e^{-iH\delta t} \bar{\pi} e^{-iH\delta t} \cdots \bar{\pi} e^{-iH\delta t} 
\nonumber \\
    & \simeq & \bar{\pi}(1-iH\delta t) \bar{\pi}(1-iH\delta t) \cdots \bar{\pi}(1-iH\delta t) 
\nonumber \\ 
    & = & \bar{\pi}(1-iH\delta t) \bar{\pi}\bar{\pi} (1-iH\delta t) \cdots \bar{\pi}\bar{\pi}(1-iH\delta t) ,
\nonumber \\
\end{eqnarray}
where we have used $\bar{\pi}=\bar{\pi}^2$. We assume that the initial state $|\psi_0\rangle$ 
lies in the Hilbert space $\overline{\cal H}=\bar{\pi}{\cal H}$, i.e. that 
\begin{equation}
|\psi_0\rangle=\bar{\pi}|\psi_0\rangle .
\end{equation}
Hence we can write 
\begin{eqnarray}\label{V^k} 
 V^k|\psi_0\rangle & \simeq & 
\bar{\pi}(1-iH\delta t) \bar{\pi}\cdot \bar{\pi} (1-iH\delta t) \bar{\pi} 
\nonumber \\
& & \cdots \bar{\pi}(1-iH\delta t) \bar{\pi}|\psi_0\rangle 
\nonumber \\
& = & (\bar{\pi}-i\overline{H}\delta t)(\bar{\pi}-i\overline{H}\delta t)
\cdots (\bar{\pi}-i\overline{H}\delta t)|\psi_0\rangle
\nonumber \\
& = & (1-i\overline{H}\delta t) (1-i\overline{H}\delta t) \cdots (1-i\overline{H}\delta t) |\psi_0\rangle
\nonumber \\
& = & (1-i\overline{H}\delta t)^k |\psi_0\rangle ,
\end{eqnarray}
where
\begin{equation}\label{barH}
 \overline{H}\equiv \bar{\pi}H\bar{\pi} .
\end{equation}
Finally, defining $t=k\delta t$ and assuming that $k\gg 1$
(so that $t$ is not small even though $\delta t$ is small), we can use the formula
\begin{equation}
 \lim_{k\to\infty} (1-i\overline{H}t/k)^k= e^{-i\overline{H}t}
\end{equation}
to conclude that (\ref{V^k}) can be approximated with
\begin{equation}
 V^k|\psi_0\rangle \simeq e^{-i\overline{H}t} |\psi_0\rangle .
\end{equation}
Since $\overline{H}$ is a hermitian operator, $e^{-i\overline{H}t}$ is unitary.
Hence (\ref{colV^k}) can be approximated with
\begin{equation}\label{colV^k2}
 |\psi_0\rangle \rightarrow e^{-i\overline{H}t} |\psi_0\rangle .
\end{equation}
Eq.~(\ref{colV^k2}) is quite remarkable; it tells us that the evolution governed by $H$ and 
interrupted with a large number $k$ of discrete non-unitary collapses can be approximated 
with a {\em continuous unitary} evolution governed by a modified Hamiltonian $\overline{H}$.
To compare it with (\ref{Hevol}), the content of (\ref{colV^k2}) can be written as the evolution 
\begin{equation}\label{c} 
 |\bar{\psi}_c(t)\rangle \simeq e^{-i\overline{H}t}|\psi_0\rangle ,
\end{equation}
where the label $c$ indicates that it is the {\em conditional} state, namely the state
valid for the case when no detection has happened up to time $t$.  
The bar in $|\bar{\psi}_c(t)\rangle$ indicates that this state lies in the Hilbert space
$\overline{\cal H}=\bar{\pi}{\cal H}$, which is a subspace of the full ${\cal H}$.

Now what if the detection has not happened up to time $t-\delta t$ but finally happened 
at the time $t$? Then the state at time $t$ is 
$\pi|\psi_c(t)\rangle/|\pi|\psi_c(t)\rangle|$, where
\begin{equation}\label{c2}
 |\psi_c(t)\rangle \equiv e^{-iH\delta t}|\bar{\psi}_c(t-\delta t)\rangle .
\end{equation}
So {\em if} the detection has not happened up to time $t-\delta t$, {\em then} the 
conditional probability that the detection will happen at time $t$ is 
\begin{equation}\label{p} 
 p(t)=\langle\psi_c(t)|\pi|\psi_c(t)\rangle .
\end{equation}
This can also be written as
\begin{equation}\label{p'} 
 p(t)=\langle\bar{\psi}_c(t-\delta t)|e^{iH\delta t}\pi e^{-iH\delta t}|\bar{\psi}_c(t-\delta t)\rangle ,
\end{equation}
which shows that in the limit $\delta t\to 0$ we have
\begin{equation}\label{p2}
 p(t)\stackrel{\delta t\to 0}{\longrightarrow} \langle\bar{\psi}_c(t)|\pi|\bar{\psi}_c(t)\rangle =0.
\end{equation}

But what is the probability that the detection will not happen up to the time $t-\delta t$?
This is the probability $1-p(t_1)$ that the detection will not happen at $t_1=\delta t$, times 
the probability $1-p(t_2)$ that the detection will not happen at $t_2=2\delta t$ (given that 
it has not happened at $t_1$), ...,
times the probability $1-p(t_{k-1})$ that the detection will not happen at $t_{k-1}=(k-1)\delta t$ (given
that it has not happened at $t_{k-2}$).
Hence the overall probability that the detection will happen at time $t=t_k$ is
\begin{eqnarray}\label{P}
 P(t) & = & p(t_k)[1-p(t_{k-1})]\cdots[1-p(t_2)][1-p(t_1)] 
\nonumber \\
& = & p(t_k) e^{ \sum_{i=1}^{k-1} \ln[1-p(t_i)] }
\nonumber \\
& \simeq & p(t_k) e^{ -\sum_{i=1}^{k-1} p(t_i) } ,
\end{eqnarray}
where in the last line we have used the approximation $\ln[1-p(t_i)]\simeq -p(t_i)$, valid because
(\ref{p2}) shows that $p(t_i)$ is a small quantity for small $\delta t$. 
For small $\delta t$ we can approximate the sum with the integral 
\begin{equation}
 \sum_{i=1}^{k-1}=\sum_{i=1}^{k-1} \frac{\delta t}{\delta t} \simeq \int_0^{t-\delta t} \frac{dt}{\delta t}
\simeq \int_0^{t} \frac{dt}{\delta t} ,
\end{equation}
so (\ref{P}) can be written in the final form as
\begin{equation}\label{Pdens}
 {\cal P}(t) \simeq w(t) e^{-\int_0^t dt'w(t')} ,
\end{equation}
where 
\begin{equation}\label{Pw}
 {\cal P}(t)\equiv\frac{P(t)}{\delta t}, \;\;\; w(t)\equiv\frac{p(t)}{\delta t} 
\end{equation}
are probability {\em densities}. 

Eq.~(\ref{Pdens}) is our main final result, with $w(t)$ being defined through 
Eqs.~(\ref{Pw}), (\ref{p}), (\ref{c2}) and (\ref{c}). It coincides with Eqs.~(\ref{I2})-(\ref{I4})
in the Introduction.

Finally, let us briefly present an alternative derivation of (\ref{Pdens}), by a reasoning that can be viewed as 
complementary to (\ref{P}). We want to determine the probability $P(t_k)$ that the event will be detected at the time $t_k$.
The probability that the event will be detected up to the time $t_{k-1}$ is
$\sum_{i=1}^{k-1} P(t_i)$, so  
the probability that it will {\em not} be detected up to the time $t_{k-1}$
is $1-\sum_{i=1}^{k-1} P(t_i)$. Therefore, instead of the first line in (\ref{P}),
alternatively we can write
\begin{equation}\label{alt1}
 P(t)=p(t_k) \left[ 1-\sum_{i=1}^{k-1} P(t_i) \right] .
\end{equation}
Approximating the sum with the integral and defining ${\cal P}(t)=P(t)/\delta t$, $w(t)=p(t)/\delta t$ as before, 
it becomes an integral equation
\begin{equation}\label{alt2}
 {\cal P}(t) \simeq w(t) \left[ 1- \int_0^{t} dt'\, {\cal P}(t') \right] .
\end{equation}
To transform the integral equation into a differential one, we take the time derivative of (\ref{alt2})
\begin{equation}\label{alt3}
 \dot{\cal P}\simeq \dot{w}\left[ 1- \int_0^{t} dt'\, {\cal P}(t') \right] -w{\cal P}
\simeq \dot{w}\frac{\cal P}{w} -w{\cal P},
\end{equation}
where the dot denotes the time derivative. The resulting differential equation can then be written as
\begin{equation}\label{alt4}
 \frac{d{\cal P}}{\cal P} \simeq \left[\frac{d\ln w}{dt} -w \right] dt,
\end{equation}
which is easily integrated to yield (\ref{Pdens}).

\section{General notes}
\label{SECnotes}

\subsection{Relation to related work}

Our approach, based on a series of quantum jumps in Eqs. (\ref{collapse1})-(\ref{colV^k}), 
can be thought of as a version of a larger class of approaches such as quantum trajectory 
approach \cite{carmichael1,carmichael2}, quantum jump approach \cite{hegerfeldt,plenio},
Monte Carlo wave-function approach \cite{dalibard} and nonlinear diffusion approach \cite{gisin}.  
(The review \cite{plenio} contains also a discussion of other similar approaches.)
The essential novelty of our approach, however, lies in the analysis in (\ref{V^k_first})-(\ref{c}). 
In particular, in our approach the conditional evolution is given by a {\em hermitian}
Hamiltonian $\overline{H}$ defined by (\ref{barH}), while in comparable approaches 
\cite{hegerfeldt,plenio,dalibard} the conditional evolution is given by a {\em non-hermitian} effective Hamiltonian.
Furthermore, the jumps in \cite{carmichael2} are modeled by creation and destruction operators,
rather than with projectors.
Besides, the waiting time studied in \cite{carmichael1,carmichael2} is not the same thing as 
awaiting time in our approach. While the waiting time in \cite{carmichael1,carmichael2} is the time
between {\em two} detections of {\em two} photons, thus giving information about statistics in many-photon states,
our awaiting time refers to {\em one} detection, most interesting in the case of a {\em one} particle state.   
Finally, the approach in \cite{gisin} is based on a Lindblad equation with an additional stochastic term,
while our approach is not based on a Lindblad equation and does not contain a stochastic term.

\subsection{Can we let $\delta t\to 0$?}

The results in the previous section were obtained in the approximation of sufficiently small $\delta t$.
Can we just take the limit $\delta t\to 0$ and say that the equations are exact in that limit?
The answer is that we cannot. Not because the limit wouldn't exist (mathematically it exists!), 
but because the limit would be trivial and physically uninteresting. This is seen as follows.
By expanding $e^{iH\delta t}$ and $e^{-iH\delta t}$ in (\ref{p'}) into powers of $\delta t$
and using $\langle \bar{\psi}_c(t-\delta t)|\pi=0$, $\pi|\bar{\psi}_c(t-\delta t)\rangle=0$,
one finds that the terms proportional to $(\delta t)^0$ and $(\delta t)^1$ do not contribute
and that the lowest non-vanishing contribution is
\begin{equation}\label{p2zeno}
p(t)=(\delta t)^2\langle \bar{\psi}_c(t-\delta t)|H\pi H|\bar{\psi}_c(t-\delta t)\rangle +{\cal O}((\delta t)^4).
\end{equation}
Note that it is {\em quadratic} in $\delta t$, rather than linear.
The consequence is that the corresponding probability density $w(t)$ in (\ref{Pw}) vanishes 
in the limit $\delta t\to 0$, in which case (\ref{Pdens}) reduces to the trivial result 
\begin{equation}
 {\cal P}(t)=0 .
\end{equation} 
In other words, if one checks with infinite frequency whether an event has happened,
then the event will never happen. This seemingly paradoxical result is in fact well known
as the quantum Zeno effect \cite{misra,decoh1,zeno-review,auletta}.

\subsection{Total probability}

What is the probability that the detection will eventually happen at {\em any} time?
It is simply $\int_0^\infty dt\, {\cal P}(t)$. Introducing a new variable
\begin{equation}
 u(t)=\int_0^t dt'w(t') ,
\end{equation}
from (\ref{Pdens}) we see that
\begin{equation}
 \int_0^\infty dt\, {\cal P}(t)=\int_{u(0)=0}^{u(\infty)} du' e^{-u'}=1-e^{-u(\infty)}\le 1 .
\end{equation}
In particular, the detection will sooner or later happen with certainty if and only if $u(\infty)=\infty$.

Note that $\int_0^{\infty} dt\, w(t)$ can be larger than 1, 
which is consistent because $w(t)$ is a probability ``density'' in a different sense than
${\cal P}(t)$. While different $t$'s in ${\cal P}(t)$ label different random events,
different $t$'s in $w(t)$ label different conditions under which an event happens.
Let us illustrate it by an example in classical probability. Suppose that a player plays roulette
every minute, each time putting money on a single number. Then $\delta t=1$ minute, each time the probability 
of winning is $p=\frac{1}{37}$ and 
$w(t)=\frac{p}{\delta t}=\frac{1}{37}$ per minute is $t$-independent. Hence $\int_0^t dt'\,w(t')=wt$, 
which is larger than 1 when $t>37$ minutes. 
Now suppose that the player decides to play until he wins. Then (\ref{I2}) gives the 
probability density that he will win at the time $t$, so the average time needed for winning is 
$\langle t \rangle=\int_0^{\infty} dt\, t {\cal P}(t) = \int_0^{\infty} dt\, tw e^{-wt}=\frac{1}{w}=37$ minutes.

\subsection{POVM} 

The most general measurements in QM can be described as POVM measurements 
\cite{peres,nielsen-chuang,muynck,audretsch,schumacher,laloe,witten}.
The measurement of the time of detection as described in this paper is not an exception.
Eqs.~(\ref{I2})-(\ref{I4}) can be written as 
\begin{equation}
 {\cal P}(t)=\langle\psi_0|{\cal E}(t)|\psi_0\rangle ,
\end{equation}
where ${\cal E}(t)$ is a POVM operator
\begin{equation}\label{POVM}
 {\cal E}(t)={\cal K}^{\dagger}(t){\cal K}(t) ,
\end{equation}
\begin{equation}\label{POVMK}
 {\cal K}(t)=\frac{e^{-\int_0^t dt'w(t')/2}}{\sqrt{\delta t}} 
\pi e^{-iH\delta t} e^{-i\overline{H}(t-\delta t)}  
\end{equation}
(and we still use units $\hbar=1$.)
Those operators, together with 
$\overline{E}\equiv 1-\int_0^{\infty} dt\, {\cal E}(t)$,  
make a resolution of the identity
\begin{equation}
 \overline{E}+\int_0^{\infty} dt\, {\cal E}(t) =1.
\end{equation}

A peculiar (yet consistent) feature of the POVM operator (\ref{POVM}) is that it depends on the state 
$|\psi_0\rangle$, through the dependence on $w(t')$ in (\ref{POVMK}) which depends on $|\psi_0\rangle$
by (\ref{I3})-(\ref{I4}).

\begin{figure*}[t]
\centering
\includegraphics[width=12cm]{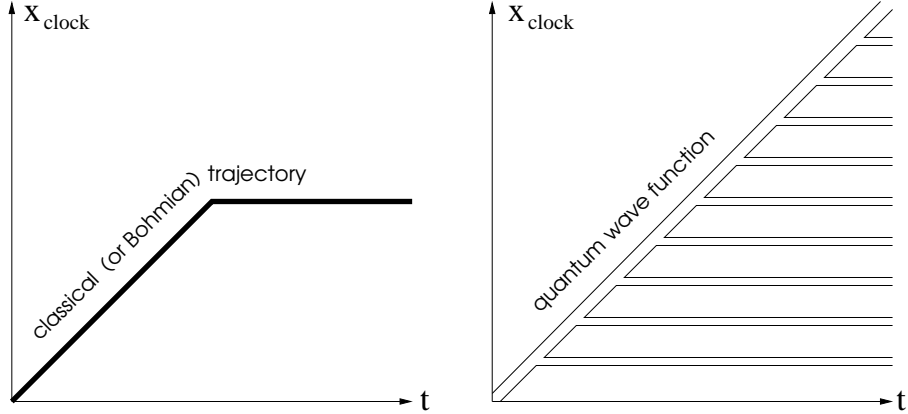}
\caption{\label{fig1}
{\it Left:} The stopwatch modeled as a clock with a needle at the position $x_{\rm clock}$ as a function of time $t$.
Initially the clock is running with $x_{\rm clock}=vt$, but at certain time the clock stops, after which 
$x_{\rm clock}$ does not longer change with time. The trajectory $x_{\rm clock}(t)$ can be thought of
as a classical trajectory, or, alternatively, as a (coarse grained) trajectory of clock particles 
in the Bohmian interpretation.
{\it Right:} In the unitary quantum description of the stopwatch, the wave function of the clock has many branches, 
corresponding to many different positions $x_{\rm clock}$ at which the clock may stop.   
}
\end{figure*}
\begin{figure*}[h]
\centering
\includegraphics[width=14cm]{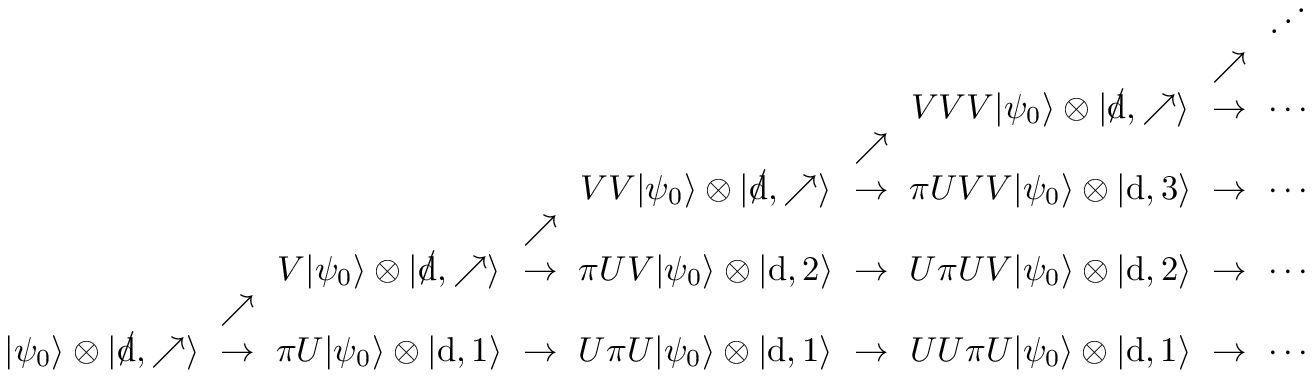}
\caption{\label{fig2}
The branching structure of the unitary evolution in the Hilbert space ${\cal H}\otimes{\cal H}_{\rm apparatus}$,
where the apparatus consists of the detector and the stopwatch. In ${\cal H}$, the evolution from the initial state
$|\psi_0\rangle$ is expressed in terms of the intrinsic Hamiltonian $H$ and the projectors $\pi$ and $\bar{\pi}=1-\pi$,
with the notation $U=e^{-iH\delta t}$ and $V=\bar{\pi}U$. In the apparatus state $|\slashed{\rm d},\nearrow\rangle$,
``$\slashed{\rm d}$'' denotes that the detector has not detected an event, while ``$\nearrow$'' denotes that the clock 
is running. In the apparatus state $|{\rm d},1\rangle$, ``${\rm d}$'' denotes that the detector has detected an event,
while ``1'' denotes that the needle of the clock stopped at the position $x_{\rm clock}=x_1$. Similarly 
in  $|{\rm d},2\rangle$, ``2'' denotes that the needle of the clock stopped at $x_{\rm clock}=x_2$, etc.  
}
\end{figure*}

\subsection{Unitarity, branching and Bohmian mechanics}
\label{SECunitary}

So far we formulated the theory in terms of ``wave function collapses'' induced by measurements.
But measurement can also be described in a fully unitary manner, without an explicit collapse,
provided that the quantum state of the measuring apparatus is also taken into account.
Such a formulation is particularly important in the theory of decoherence \cite{decoh1,decoh2},
as well as in the many world \cite{mw1,mw2,mw3} and Bohmian \cite{bohm,book-bohm,book-hol,book-durr,oriols}
interpretations of QM. In the unitary description, every measurement with more than one possible outcomes
is associated with a {\em branching} of the full quantum state (describing the measured system and the apparatus),
with one branch for each possible outcome. 

Figs.~\ref{fig1} and \ref{fig2} show how such a branching looks like 
for the theory developed Sec.~\ref{SECmain}, with possible detections at discrete times $t_1=\delta t$, 
$t_2=2\delta t$, etc. Fig.~\ref{fig1} depicts the branching of the clock wave function, while 
Fig.~\ref{fig2} shows a branching diagram for the full quantum state. Fig.~\ref{fig2} 
shows that, at time $t_i$, the full state is a superposition of $i+1$ terms, namely   
\begin{eqnarray}\label{branch}
|\Psi(t_0)\rangle & = & |\psi_0\rangle \otimes|\slashed{\rm d},\nearrow\rangle ,
\nonumber \\ 
|\Psi(t_1)\rangle & = & V |\psi_0\rangle \otimes|\slashed{\rm d},\nearrow\rangle +
\pi U|\psi_0\rangle\otimes|{\rm d},1\rangle ,
\nonumber \\
|\Psi(t_2)\rangle & = & VV |\psi_0\rangle\otimes |\slashed{\rm d},\nearrow\rangle +
\pi UV |\psi_0\rangle\otimes|{\rm d},2\rangle +
\nonumber \\
 & & U\pi U|\psi_0\rangle\otimes|{\rm d},1\rangle ,
\nonumber \\
|\Psi(t_3)\rangle & = & VVV|\psi_0\rangle\otimes|\slashed{\rm d},\nearrow\rangle + \pi UVV |\psi_0\rangle\otimes|{\rm d},3\rangle +
\nonumber \\ 
& & U\pi UV |\psi_0\rangle\otimes|{\rm d},2\rangle + UU\pi U|\psi_0\rangle\otimes|{\rm d},1\rangle ,
\nonumber \\
 & \vdots &
\end{eqnarray}

In particular, Fig.~\ref{fig1} can be used to understand why is the formula (\ref{P}) for
probability $P(t_k)$ of detection at time $t_k$, derived with standard QM, valid also
in the Bohmian interpretation. Essentially, this is because
Fig.~\ref{fig1} shows how the measurement of time is reduced to a measurement 
of the {\em position} of something \cite{nik-ibm}, 
while probabilities of positions in the Bohmian interpretation are the same as probabilities of positions 
in the standard QM \cite{bohm,book-bohm,book-hol,book-durr,oriols,nik-ibm}.

It should also be noted that in the literature \cite{leavens,durrtime1,durrtime2}
the arrival time has been calculated with Bohmian mechanics in a different way, 
without taking into account the behavior of the measuring apparatus. 
Such a calculation may give a time distribution of particle arrivals that differs from that in our theory
(see Sec.~\ref{SECarrival}).
However, a result obtained without taking into account the behavior of the measuring apparatus
is not directly relevant for making measurable predictions. In general, 
Bohmian mechanics makes the same measurable predictions as standard QM only when the 
behavior of the measuring apparatus is taken into account, as in Figs.~\ref{fig1} and \ref{fig2} 
and Eq.~(\ref{branch}).

\section{Decay of an unstable state}
\label{SECdecay}

\subsection{Decay from the general theory}

Now we want to understand how the general formalism developed in Sec.~\ref{SECmain}
can be applied to study a decay of an unstable system. 
Hence we assume that the initial state $|\psi_0\rangle$ is an unstable state that can decay into many different 
states, called decay states, orthogonal to $|\psi_0\rangle$. Furthermore, we assume that the detector 
(which may be comprised of many small detectors) can detect {\em any} of those decay states.
Hence we can take
\begin{equation}
 \pi=1-|\psi_0\rangle\langle\psi_0| ,
\end{equation}
so $\bar{\pi}=|\psi_0\rangle\langle\psi_0|$ and the evolution governed by $\overline{H}$ just keeps the state
in the initial state $|\psi_0\rangle$, up to an irrelevant time-dependent phase. Therefore 
we can write $|\bar{\psi}_c(t)\rangle=|\psi_0\rangle$ and (\ref{p'}) reduces to
\begin{eqnarray}\label{p'd} 
 p & = & \langle\psi_0|e^{iH\delta t}\left[1-|\psi_0\rangle\langle\psi_0|\right]e^{-iH\delta t}|\psi_0\rangle
\nonumber \\
& = & 1 -|\langle\psi_0|e^{-iH\delta t}|\psi_0\rangle|^2 ,
\end{eqnarray}
where we have used the normalization $\langle\psi_0|\psi_0\rangle=1$. 

\subsection{Decay for very small $\delta t$}

By expanding $e^{-iH\delta t}$ in (\ref{p'd})
up to the term quadratic in $\delta t$ and restoring the units $\hbar\neq 1$, we obtain 
\begin{equation}\label{zeno2}
 p \simeq \frac{(\delta t)^2}{\hbar^2} [\langle\psi_0|H^2|\psi_0\rangle -\langle\psi_0|H|\psi_0\rangle^2]
= \frac{(\delta t)^2 (\Delta H)^2}{\hbar^2} ,
\end{equation}
where $\Delta H$ is the uncertainty of energy defined by $H$, in the state $|\psi_0\rangle$.
Eq.~(\ref{zeno2}) is just a special case of (\ref{p2zeno}) and is a well-known result 
in the study of the quantum Zeno effect for decays \cite{decoh1,zeno-review,auletta}.

\subsection{Quasi-spontaneous decay}

What is usually called a ``spontaneous'' decay in the literature is in fact a quasi-spontaneous decay.
One often thinks of a decay as a process in which a micro system, say an atom, randomly 
jumps into a more stable state. Intuitively, one often imagines that this jump 
is spontaneous, in the sense that nothing outside of the micro system influences it. 
But in standard QM, a random jump is in fact a ``wave function collapse'' 
induced by some kind of ``measurement'', where ``measurement'' always involves decoherence induced by a large number of 
environment degrees of freedom \cite{decoh1,decoh2}. So there can be no random jump without environment.
According to standard QM, in a hypothetic universe containing only one excited atom and nothing else, 
a random jump should never happen. In this sense there is no such thing as spontaneous decay.
At best we can have a decay which does not depend on {\em details} of the 
environment, creating an illusion that the environment is not important at all, which we refer to as 
a quasi-spontaneous decay.

Let us briefly explain how quasi-spontaneous decay can be understood within our theory.
The only quantitative property of the environment in our theory is the time resolution $\delta t$
of the detector. Hence we can say that the decay is quasi-spontaneous when $w$ in (\ref{Pdens}) does not depend on
$\delta t$. From (\ref{Pw}) we see that this means that $p/\delta t$ does not depend on $\delta t$, 
where $p$ is given by (\ref{p'd}). Therefore the decay is quasi-spontaneous when (\ref{p'd}) is
proportional to $\delta t$, i.e. when we can write
\begin{equation}\label{gammat}
 p=\Gamma\delta t .
\end{equation}
In this case we have $w=\Gamma$, so (\ref{I2}) reduces to
\begin{equation}
 {\cal P}(t)=\Gamma e^{-\Gamma t} .
\end{equation}
Hence the probability that the decay will not happen at time $t$ or before is
\begin{equation}\label{exp}
 \overline{P}(t)=1-\int_0^t dt'\, {\cal P}(t')= e^{-\Gamma t}, 
\end{equation}
which is nothing but the usual exponential law for the survival probability. 

But how can the linear law (\ref{gammat}) be valid without contradicting (\ref{zeno2})? 
The answer is that (\ref{zeno2}) is applicable to very short times $\delta t$,
while the linear law is an approximation applicable to larger times $\delta t$.
This can be seen from the literature \cite{fonda,giacosa,pascazio} 
where the survival probability at time $t$ is computed without assuming measurements
before $t$, which in our formulation is equivalent to considering the case $\delta t=t$.
The computations in \cite{fonda,giacosa,pascazio} show that the exponential law 
(\ref{exp}) is just an approximation, approximately valid for a large range of intermediate times $t$,
but completely wrong for very short and very long times. In our formulation this means that
$\overline{P}(\delta t)\approx e^{-\Gamma \delta t}$ for intermediate $\delta t$'s, which corresponds
to the approximate linear law (\ref{gammat}) for intermediate $\delta t$'s.  

\section{Arrival time}
\label{SECarrival}

Suppose that a wide 1-particle wave packet travels towards a detector laying in the $x$-$y$ plane at $z=0$.
One is interested in the probability density that the particle will arrive to the detector
at the time $t$. The corresponding measurable quantity is the probability density that the particle will 
be {\em detected} at the time $t$.
The corresponding projector can be taken to be 
\begin{equation}
 \pi = \int_{-\infty}^{\infty} dx \int_{-\infty}^{\infty} dy \int_{-l/2}^{l/2} dz\,
|x,y,z\rangle\langle x,y,z| , 
\end{equation}
where $l$ is the width of the detector in the $z$-direction. Eq.~(\ref{I3}) can then be written as 
\begin{equation}
 w(t)=\frac{1}{\delta t} \int_{-\infty}^{\infty} dx \int_{-\infty}^{\infty} dy \int_{-l/2}^{l/2} dz\,
|\psi_c(x,y,z,t)|^2 ,
\end{equation}
where $\psi_c(x,y,z,t)=\langle x,y,z|\psi_c(t)\rangle$. By taking $H$ to be the free Hamiltonian
$H={\bf p}^2/2m$, the rest of the analysis is, in principle, straightforward 
(but possibly complicated in practice).

There is, however, one conceptual issue that we want to resolve. The exponential factor in (\ref{I2})
seems to suggest that ${\cal P}(t)$ decreases with time. While such a decrease is something to be expected 
in the case of a decay (Sec.~\ref{SECdecay}), it is not expected in the case of an arrival time.
So what is the physical meaning of the exponential factor in (\ref{I2})?

\begin{figure}[t]
\centering
\includegraphics[width=5cm]{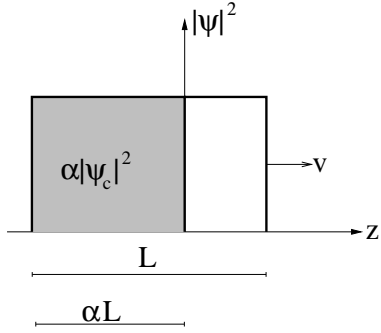}
\caption{\label{fig3}
The rectangular wave packet $|\psi(z)|^2$ of width $L$  at a time $t$, for $0<t<T$. 
The packet moves to the right with the velocity $v$. The corresponding 
conditional wave packet $|\psi_c(z)|^2$ (gray region) has support only for $z<0$, so its width is
$\alpha L$ with $\alpha=(T-t)/T$. The correct normalization of the conditional wave packet 
gives $|\psi_c(z)|^2=\alpha^{-1}|\psi(z)|^2$ for $z<0$, so in this region $|\psi(z)|^2=\alpha|\psi_c(z)|^2$. 
}
\end{figure}

We stress that there is nothing inherently quantum about the exponential factor.
It arises from the classical multiplication of probabilities in (\ref{P}). 
Hence, to understand qualitatively the meaning of the exponential factor in the time of arrival,
it is illuminating to study a purely classical setup. So consider a 1-dimensional system 
in which a classical particle moves with the constant velocity $v$
from $z<0$ towards the detector at $z=0$. We consider a classical statistical ensemble in which the velocity is known exactly, 
while the knowledge of the position is described by a classical probability density $\rho(z,t)$.
The classical analog of the wave function is then simply $\psi(z,t)=\sqrt{\rho(z,t)}$.
To further simplify the analysis, we assume that $\delta t\to 0$ (which is allowed in the classical setting)
and that $\rho(z,t)$ is a rectangular distribution for fixed $t$. Thus we think of $\psi(z,t)$ 
as a 1-dimensional rectangular wave packet (Fig.~\ref{fig3}), with $|\psi(z,t)|^2=1/L$ inside the packet and 
$|\psi(z,t)|^2=0$ outside of the packet. 
The packet moves with the velocity $v$ from the region $z<0$ towards the detector at $z=0$. 
Assuming that the front end of the packet approaches the detector at $t=0$, it follows that
the rear end approaches the detector at the time $T=L/v$. Clearly, the probability density of detecting
the particle at the time $t$ can only be non-zero for $0<t<T$. For the rectangular wave packet
one expects that this probability density should be uniform, i.e. that ${\cal P}(t)=1/T$ for $0<t<T$. 
How can that be compatible with the exponential factor in (\ref{I2})? 

To answer that question, it is essential to have in mind that (\ref{I3}) depends on the conditional state 
$|\psi_c(t)\rangle$, and not on the non-conditional state $|\psi(t)\rangle$. In our classical case,
for $t>0$, the non-conditional wave function $\psi(z,t)$ has a part that does not vanish 
for some $z>0$. But this part originates from parts of the wave function that traveled through $z=0$,
which in $\psi_c(z,t)$ are absent because the conditional probability density $\rho_c(z,t)=|\psi_c(z,t)|^2$
is the probability density conditioned on the assumption that the particle has not been detected at $z=0$
before the time $t$, so there can be no particle in the region $z>0$.
(In the quantum case, formally, those parts of the wave 
function are absent because they are removed by a series of $V$-operators that involve the projections $\bar{\pi}$.) 
Hence $|\psi_c(z,t)|^2=0$ for $z>0$ and $|\psi_c(z,t)|^2=[T/(T-t)] |\psi(z,t)|^2$ for $z<0$,
as illustrated by Fig.~\ref{fig3}. 
So given that the particle has not been detected before $t$, the probability density that 
it will be detected at time $t$ is
\begin{equation}
 w(t)=\left\{ 
\begin{array}{ll}
 \displaystyle\frac{1}{T-t} & \;\; {\rm for} \;\; 0<t<T  \\
 0 & \;\; {\rm otherwise} .
\end{array}
\right.
\end{equation}
Hence
\begin{equation}
 u(t)=\int_0^t dt'\, w(t') = \left\{ 
\begin{array}{ll}
 \ln\displaystyle\frac{T}{T-t} & \;\; {\rm for} \;\; 0<t<T  \\
 \infty &  \;\; {\rm for} \;\; t\geq T ,
\end{array}
\right.
\end{equation}
so (\ref{I2}) finally gives
\begin{equation}\label{noexp}
 {\cal P}(t)=\frac{1}{T-t} e^{-\ln\frac{T}{T-t} }=\frac{1}{T} 
\end{equation}
for $0<t<T$, as expected. Eq.~(\ref{noexp}) represents the simplest demonstration
of how the exponential factor in (\ref{I2}) may give raise to a 
probability density that does not decrease with time $t$. 
  
\section{Dwell time and tunneling time}
\label{SECdwelltunn}

In the case of dwell time one asks how much time a particle will spend
in a spatial region $V$, starting from $t=0$ and assuming that the particle is in $V$ initially at $t=0$. 
Operationally we can say that the particle is in the region 
$V$ until it gets detected in its complement $\overline{V}$. 
Hence the relevant projector is 
\begin{equation}
 \pi=\int_{\overline{V}} d^3x\, |{\bf x}\rangle\langle{\bf x}| .
\end{equation}
The probability that the particle will leave $V$ at time $t$ is the probability that it will be detected 
in $\overline{V}$ at time $t$. Hence the average time needed to leave $V$, that is the average dwell time, is
\begin{equation}\label{taver}
 \langle t\rangle = \int_0^{\infty} dt\, t{\cal P}(t),
\end{equation}
where ${\cal P}(t)$ is given by (\ref{I2}).

In the case of tunneling, the full space consists of one classically forbidden region $V_{\rm forb}$
and two classically allowed regions $V_1$ and $V_2$. 
In the tunneling time problem, one asks how much time the particle will spend in $V_{\rm forb}$, 
given that initially the particle is in $V_1$. Assuming that there is no detector in $V_{\rm forb}$,
the problem can be solved by having two detectors, one in $V_1$ and the other in  $V_2$.  
One first uses a detector in $V_1$ to frequently check whether the particle is still in $V_1$. 
When, at a certain time, it happens that the particle is no longer in $V_1$, we call this time
$t_0=0$ and study the response of the second detector in $V_2$. The relevant projector 
associated with the second detector is
\begin{equation}
 \pi=\int_{V_2} d^3x\, |{\bf x}\rangle\langle{\bf x}| .
\end{equation}
Hence ${\cal P}(t)$ given by (\ref{I2}) is the probability density that the particle will be detected 
in $V_2$ at time $t$. Between $t_0=0$ and $t$ the particle can be considered to be in  
$V_{\rm forb}$, so the average time spent there is again given by a formula of the form (\ref{taver}).

\section{Conclusion}
\label{SECconclusion}

In this paper we have developed a general theory of computing probability density
that an awaited event will be detected at the time $t$, in the approximation in which 
the time resolution of the detector $\delta t$ is sufficiently small, so that 
the quantum state of the measured system does not change much during $\delta t$ by the evolution governed by the intrinsic Hamiltonian $H$ of the measured system. The theory does not depend on any details of the detector, except on its time resolution 
$\delta t$. The awaited event is defined by a projector $\pi$ in the Hilbert space 
of the measured system. Time is treated as a classical parameter and no ``time operator'' is needed. The measurement of time is reduced to an observation of a position of a macroscopic pointer, such as the position of the needle of a clock.
The theory is based on the usual ``collapse'' postulate 
induced by quantum measurements, but the predictions of the theory do not depend
on whether the ``collapse'' is interpreted as a real physical event or merely as an
information update. In particular, the predictions of the theory are also consistent with the many world and the Bohmian interpretation, in which no real collapse is present.

Being concentrated on general theory, in this paper we have outlined the general 
principles of how the theory can be applied to some more specific problems (decay, arrival, dwell and tunneling time), but we 
have not analyzed any such realistic problem in detail. Concrete applications of the theory are left as a project for the future work.   

\section*{Acknowledgments}

This work was supported  
by the European Union through the European Regional Development Fund 
- the Competitiveness and Cohesion Operational Programme (KK.01.1.1.06).

\end{document}